\documentclass[twocolumn,showpacs,preprintnumbers,amsmath,amssymb]{revtex4}
\usepackage{graphicx}
\usepackage{dcolumn}
\usepackage{bm}
\usepackage{mathrsfs}

\begin{document}
\preprint{APS/123-QED}

\title{Core-coupled states and split proton-neutron quasi-particle multiplets in $^{122-126}$Ag}

\author{S.~Lalkovski$^{1,2}$, A.~M.~Bruce$^2$, A.~Jungclaus$^3$, M.~G\'orska$^4$, M.~Pf\"utzner$^5$, 
L.~C\'aceres$^{4,6}$\footnote{Present address: GANIL CAEN France}, F.~Naqvi$^{4,7}$, 
S.~Pietri$^4$, Zs.~Podoly\'ak$^{8}$, G.S.~Simpson$^{9}$, K.Andgren$^{10}$, 
P.~Bednarczyk$^{4,11}$, T.~Beck$^4$, J.~Benlliure$^{12}$, G.~Benzoni$^{13}$, E.~Casarejos$^{14}$, 
B.~Cederwall$^{10}$, F.~C.~L.~Crespi$^{13,15}$, J.~J.~Cuenca-Garc\'ia$^4$, I.~J.~Cullen$^{8}$,  
A.~M.~Denis~Bacelar$^2$, P.~Detistov$^{16}$, P.~Doornenbal$^{4,7}$, G.~F.~Farrelly$^{8}$, 
A.~B.~Garnsworthy$^{8}$, H.~Geissel$^4$, W.~Gelletly$^{8}$, J.~Gerl$^4$, J.~Grebosz$^{4,11}$, 
B.~Hadinia$^{10}$, M.~Hellstr\"om$^{17}$, C.~Hinke$^{18}$, R.~Hoischen$^{17,4}$, G.~Ilie$^7$, 
G.~Jaworski$^{19,20}$, J.~Jolie$^7$, A.~Khaplanov$^{10}$, S.~Kisyov$^1$, M.~Kmiecik$^{11}$, 
I.~Kojouharov$^4$, R.~Kumar$^{21}$, N.~Kurz$^4$, A.~Maj$^{11}$, S.~Mandal$^{22}$, 
V.~Modamio$^3$\footnote{Present address: Istituto Nazionale di Fisica Nucleare, 
Laboratori Nazionali di Legnaro, Legnaro I-35020, Italy},
F.~Montes$^4$, S.~Myalski$^{11}$, M.~Palacz$^{20}$, W.~Prokopowicz$^4$, P.~Reiter$^7$, 
P.~H.~Regan$^{8}$, D.~Rudolph$^{17}$, H.~Schaffner$^4$, D.~Sohler$^{23}$, 
S.J.~Steer$^{8}$, S.~Tashenov$^4$, J.~Walker$^{3,6}$, P.M.~Walker$^{8}$, H.~Weick$^4$, 
E.~Werner-Malento$^{24}$, O.~Wieland$^{13}$, H.~J.~Wollersheim$^4$, and M.~Zhekova$^1$
}

\affiliation{
$^1$Faculty of Physics, University of Sofia ``St. Kliment Ohridski'', Sofia 1164, Bulgaria\\
$^2$School of Computing, Engineering and Mathematics, University of Brighton, Brighton BN2 4JG, UK\\
$^3$Instituto de Estructura de la Materia, CSIC, E-28006 Madrid, Spain\\
$^4$GSI Helmholtzzentrum f\"ur Schwerionenforschung, Planckstr 1, D-64291 Darmstadt, Germany\\
$^5$Faculty of Physics, University of Warsaw, PL-00681 Warsaw, Poland\\
$^6$Departamento de F\'isica Te\'orica, Universidad Autonoma de Madrid, E-28049 Madrid, Spain\\ 
$^7$Institut f\"ur Kernphysik, Universit\"at zu K\"oln, D-50937 K\"oln, Germany\\
$^8$Department of Physics, University of Surrey, Guildford GU2 7XH, United Kingdom\\
$^{9}$LPSC, Universit\'e Joseph Fourier Grenoble, CNRS/IN2P3, Institut National Polytechnique
de Grenoble, F-38026 Grenoble Cedex, France\\
$^{10}$KTH Stockholm, S-10691, Stockholm, Sweden\\
$^{11}$H.Niewodnicza\'nski Institute of Nuclear Physics, Polish Academy of Sciences,
Radzikowskiego 152 Krak\'ow 31-342, Poland\\
$^{12}$Universidad de Santiago de Compostela, E-175706 Santiago de Compostela, Spain\\
$^{13}$INFN Sezione di Milano, I-20133 Milano, Italy\\
$^{14}$Universidad de Vigo, E-36310 Vigo, Spain\\ 
$^{15}$Universit\'a degli Studi di Milano, I-20133, Milano, Italy\\
$^{16}$Institut for Nuclear Research and Nuclear Energy, Bulgarian Academy of Science, Sofia, Bulgaria\\
$^{17}$Department of Physics, Lund University, S-22100 Lund, Sweden\\
$^{18}$Physik-Department E12, Technische Universit\"at M\"unchen, D-85748 Garching, Germany\\
$^{19}$Faculty of Physics, Warsaw University of Technology, Koszykowa 75, 00-662 Warszawa, Poland\\
$^{20}$Heavy Ion Laboratory, University of Warsaw, Pasteura 5A, 02-093 Warszawa, Poland\\
$^{21}$Inter University Accelerator Centre, New Delhi, India\\
$^{22}$University of Delhi, New Delhi, India\\
$^{23}$Institute of Nuclear Research of the Hungarian Academy of Science, H-4001, Debrecen, Hungary\\
$^{24}$IEP, Warsaw University, PL-00681 Warsaw, Poland\\
}

\date{\today}

\begin{abstract}
Neutron-rich silver isotopes were populated in the fragmentation 
of a $^{136}$Xe beam and the relativistic fission of $^{238}$U. The fragments
were mass analyzed with the GSI Fragment separator and subsequently implanted
into a passive stopper. Isomeric transitions were detected by 105 HPGe 
detectors. Eight isomeric states were observed in $^{122-126}$Ag nuclei. 
The level schemes of $^{122,123,125}$Ag were revised and extended with isomeric 
transitions being observed for the first time. The excited states in the odd-mass 
silver isotopes are interpreted as core-coupled states. The isomeric states in 
the even-mass silver isotopes are discussed in the framework of the proton-neutron 
split multiplets. The results of shell-model calculations, performed for the most 
neutron-rich silver nuclei are compared to the experimental data.
\end{abstract}
\pacs{21.10.-k, 21.10.Hw, 21.10.Re, 21.10.Tg, 23.20.Lv, 23.35.+g, 27.60.+j}

\maketitle

\section{Introduction}
The Nuclear Shell model \cite{MGM48} was introduced 
in the mid-20$^{th}$ century and its major success was the description of the magic 
numbers. Recently, it was suggested that for extremely neutron-rich nuclei 
the ordering of the orbits, and hence the magic numbers as we know 
them from nuclei close to the line of $\beta$ stability, may change due to 
diffuseness of the nuclear surface \cite{Do94,SF02}. Further, it was pointed out that 
an experimental fingerprint of the changing structure may be found in the $R_{4/2}$ 
ratio systematics for the neutron-rich even-even nuclei, because in the case of 
weakening of the $\vec{l}^2$ term in the shell model potential, the $R_{4/2}$ ratio 
would strongly depend on the occupation of the $\Delta j=2$ single-particle orbits \cite{Ch95}.
The shell quenching was also suggested to be the origin of the poor 
theoretical description of the $r$-process abundance in the region below the $A=130$ peak 
\cite{Pf97, Di03}.

However, recent studies performed at GSI show no need of shell quenching to 
explain the structure of $^{130}_{\ 48}$Cd$_{82}$ \cite{Ju07}. It was pointed 
out that if a neutron shell erosion is present in the $N=82$ isotonic chain, it 
should take place deep below $^{132}$Sn in the $28\leq Z\leq 50$ proton shell. 
Among the candidates for strongly pronounced shell quenching effects are the 
neutron-rich zirconium nuclei \cite{SF02}. Of particular interest is 
$^{122}_{\ 40}$Zr$_{82}$ which would be a magic nucleus in terms of the 
``classic'' magic numbers. This nucleus is far from being accessible 
experimentally, but given that the degree of collectivity depends on the number 
of valence particles, a change in the neutron magic numbers would also affect 
the structure of the neutron mid-shell zirconium nuclei. To search for deviations 
from the mid-shell behavior, IBM-1 calculations were performed \cite{La09} 
assuming the persistence of the $N$=50 and 82 magic numbers. The spectroscopic 
observables, level energies and transition strengths in $^{106}_{\ 40}$Zr$_{66}$ 
were predicted using a Hamiltonian parametrized with respect to nuclei lying close 
to the line of $\beta$-stability. Further, the excited states in $^{106}$Zr were 
experimentally observed from $\beta$-decay studies performed at RIKEN \cite{Su11}. 
The experimental level energies are in good agreement with the model predictions 
suggesting that $^{106}$Zr$_{66}$ is indeed a mid-shell nucleus. In  $^{108}$Zr 
\cite{Su11}, an isomeric decay was also experimentally observed, revealing the 
energy of the first excited state. It was shown, that in the zirconium isotopic 
chain the $2_1^+$ state gradually increases in energy from the mid-shell nucleus 
towards the next neutron magic number. Even though the evolution of the $2_1^+$ 
states in the neutron-rich zirconium isotopes hints at the preservation of the 
``classical'' magic numbers when departing from the neutron mid-shell $^{106}$Zr 
nucleus, more experimental data are needed to draw firm conclusions.

A different approach to the problem is the systematic analysis of the states in 
the odd-mass and odd-odd nuclei and in particular the positioning of the 
unique-parity states relative to the normal-parity states. These high-$j$ 
intruder states appear at the upper part of the shells, where low-$j$ normal 
parity states are present, and are often responsible for the islands of 
isomerism emerging in the vicinity of the magic numbers. Also, in the odd-odd 
nuclei at the top of the shells, long-lived isomers emerge from the structure of 
the split proton-neutron multiplets \cite{Pa79}. Thus, the islands of isomerism 
in specific mass regions are directly related to the existence of the magic 
numbers and the identification of the opposite parity states in the odd-A and 
odd-odd nuclei can give a direct measure of whether the shells are quenched or 
not.

In the region below $^{132}$Sn, the unique-parity orbits are $\pi g_{9/2}$ and 
$\nu h_{11/2}$ arising from the fourth proton and fifth neutron oscillator shell, 
respectively, leading to the appearance of positive-parity states in the odd-A 
(Z=47) silver isotopes and states of opposite parities in the even-mass Ag nuclei. 
The interplay between these unique-parity and the normal-parity orbits often 
gives rise to isomers close to the ground state. A well-known example is 
$^{129}$Ag$_{82}$, which has a $\beta$-decaying isomer with a half-life of 
$t_{1/2}=$160 ms and a 46-ms ground state \cite{Kr00}, leading to a 
$^{129}$Ag$_{82}$ stellar half-life of the order of 80 ms. Such long-lived 
isomerism can shed light on the observed $r$-process overabundance in the 
$A\sim 120$ mass region.  Therefore, of particular interest is the search for 
isomeric states in the most neutron-rich odd-mass and odd-odd nuclei, just below 
the doubly magic $^{132}$Sn nucleus.

Prior to this study, the shell structure in nuclei around $^{132}$Sn was studied 
in a number of experiments on isomeric decays. The 
$^{125,127,129}$Sn nuclei were studied at LOHENGRIN \cite{Pi00} and during the 
$g$-RISING campaign \cite{Lo08}. Isomeric decay studies were performed for 
$^{128}$Sn \cite{Pi11}, $^{127,128,130}$Cd \cite{Ju07, Ca09, Na10} and $^{131}$In 
\cite{Go09} nuclei in the RISING Stopped beam campaign \cite{Re08} and for 
$^{123-130}$In at LOHENGRIN \cite{Sc04}. The present work extends these studies 
towards the most neutron-rich silver nuclei.

\section{Experimental Set Up}
The neutron-rich Ag nuclei were produced during two experiments performed at 
GSI, Darmstadt, using the fragmentation of a $^{136}$Xe beam 
\cite{Ju07, Ca09, Na10, Go09} and the fission of a $^{238}$U beam \cite{Br10} at 
relativistic energies. In these experiments, the beams were accelerated 
to 750 MeV/A by the SIS-18 synchrotron and impinged on Be targets of 1 and 
4 g/cm$^2$ thickness, respectively. The cocktail of fragments, produced in the 
fragmentation or fission, was analyzed with the GSI FRagment Separator (FRS)
\cite{Ge92}. The ions were separated by means of their magnetic rigidities, 
times of flight, energy losses and their positions in the middle and final focal 
plane of the separator. The nuclei were slowed by an aluminum 
wedge shaped degrader and implanted into a copper or plastic stopper, placed
at the final focal plane. Delayed $\gamma$-rays were detected 
by the RISING multidetector array \cite{Pi07}, comprising 105 HPGe 
detectors, mounted as 15 Cluster detectors. The signals were digitized 
by Digital Gamma Finder (DGF) modules providing energy and time information.

\section{Experimental Results and Data Analysis}
\begin{figure}[t]
\scalebox{0.32}[0.32]{\includegraphics{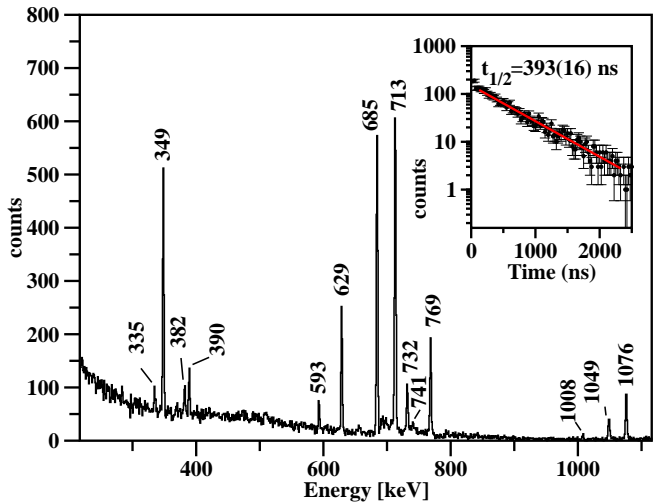}} 
\caption[]{\label{123agene} (color on-line) Gamma-rays observed in delayed 
coincidence with $^{123}$Ag ions. (inset) A summed time spectrum for the 
349, 629, 685, 713, 732 and 769-keV transitions.}
\end{figure}

\subsection{Odd-$A$ nuclei}
Prior to the current study, isomer-delayed transitions were observed in 
neutron-rich $^{123,125}$Ag \cite{St09}. Although the low-energy transitions 
directly depopulating the isomeric state were not seen in that work, half-lives of 
$t_{1/2}=396(37)$ ns and $t_{1/2}=473(111)$ ns were deduced from the $\gamma (t)$ 
distributions of the high-energy transitions de-exciting states below the isomers 
in $^{123}$Ag and $^{125}$Ag, respectively. An overall deviation of 3 keV in the 
$\gamma$-ray energies reported in \cite{St09} with respect to the present values 
is observed. Alternative level schemes from isomeric decays, were also presented 
in a PhD thesis \cite{To06}. The present work gives revised and extended level 
schemes of $^{123,125}$Ag, available due to the superior efficiency of the 
RISING multidetector array which enables $\gamma$-ray coincidences to be clearly 
established. Preliminary results on the isomeric decays in $^{123,125}$Ag  were 
reported in \cite{LA11}, but a manuscript
was not sent for the conference proceedings. During the completion of the 
present manuscript, a revised level scheme of $^{125}$Ag was published in 
\cite{Ka12}.

\subsubsection{$^{123}$Ag}
The nucleus $^{123}$Ag  was produced in both experiments and observed in the FRS 
setting centered on the transmission of fully stripped $^{120}$Rh (fission) and 
$^{126}$Cd (fragmentation) ions, respectively. 

Fig.~\ref{123agene} shows a $\gamma$-ray spectrum observed in coincidence with 
$^{123}$Ag ions, within a 3.75 $\mu$s wide coincidence window opened 125 ns after 
the implantation. The most intense transitions, presented in 
Fig.~\ref{123agene}, were previously reported and placed in level schemes 
\cite{St09, To06}. The inset of Fig.~\ref{123agene} shows a summed time spectrum 
for the strongest transitions in $^{123}$Ag. The half-life of the isomeric state 
$t_{1/2}=393$(16) ns was deduced from the fit to the slope. This value is in 
good agreement with 396(37) ns measured in \cite{St09} and within $2\sigma$ with
0.32(3) $\mu$s \cite{To06}. 

\begin{figure}[t]
\scalebox{0.34}[0.34]{\includegraphics{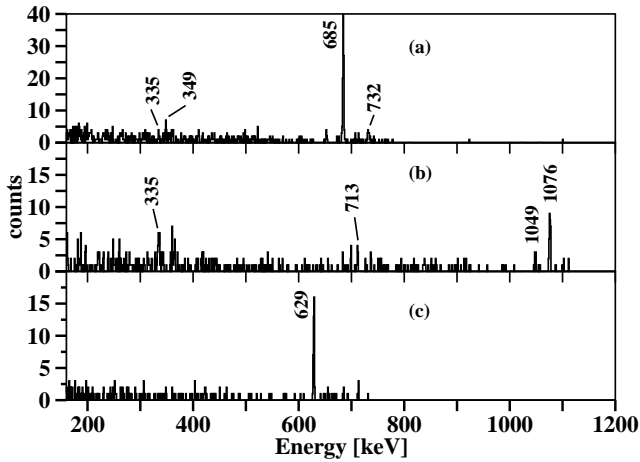}}
\caption[]{\label{7134ene}$^{123}$Ag $\gamma$-rays, observed in coincidence with 
the (a) 713-keV, (b) 349-keV and (c) 769-keV transition, respectively.}
\end{figure}

\begin{figure}[t]
\begin{center}
\rotatebox{90}{\scalebox{0.34}[0.34]{\includegraphics{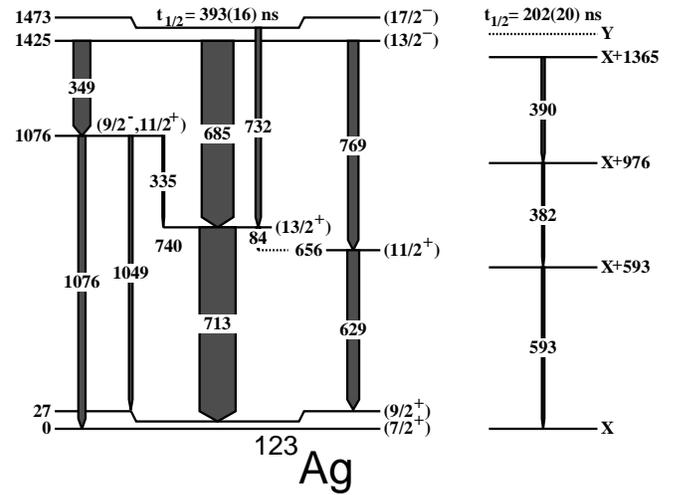}}} 
\caption[]{\label{123ag}Partial level scheme of $^{123}$Ag, based on the 
$\gamma$-ray coincidences observed in the present work.}
\end{center}
\end{figure}

Sample coincidence spectra are shown in Fig.~\ref{7134ene}. Fig.~\ref{7134ene}(a) 
shows that the 713-keV transition is in strong coincidence with the 685-keV 
transition, and in weaker coincidence with the 732-keV, 335-keV and 349-keV 
transitions. In Fig.~\ref{123agene}, the 349-keV peak is strong, while the 
335-keV peak is weak, which suggests that the 349-keV transition feeds a level, 
which subsequently decays via a branch of transitions, some of which are parallel 
to the 713-keV transition. This interpretation is supported by Fig.~\ref{7134ene}(b)
which shows the 349-keV line in coincidence with the 1049-keV and 1076-keV 
transitions. The 1049-keV and 1076-keV transitions are observed to be in mutual 
anti-coincidence. The energy spectrum, presented in Fig.~\ref{7134ene}(c) shows 
a peak with an energy of 629 keV, which is in coincidence with the 769-keV 
$\gamma$-ray. The 769-keV and 629-keV transitions are not observed to be in 
coincidence with the 713-keV and 685-keV transitions. 
\begin{table}[t]   
\begin{center}
\caption{\label{123Ag}Level energies ($E_i$), spin/parity ($J^\pi$) assignments, 
$\gamma$-ray energies ($E_\gamma$) and relative intensities for
transitions observed in coincidence with $^{123}$Ag ions.}
\begin{tabular}{cccccc} 
\hline\hline
$E_i$ (keV)&$J^\pi$&   $E_\gamma$ (keV)&$I_{rel}$ (\%) \\
\hline
       &           & 741.2(5)$^a$ & 4.3(7)\\
       &           & 1008.2(5)$^a$& 1.7(4)\\
0      & $(7/2^+)$ &          &          \\
27     &$(9/2^+)$  &          &          \\
656    & $(11/2^+)$& 629.1(5) & 32.7(15)\\
740    & $(13/2^+)$& 84(1)   & 10(5)  \\
       &           & 713.2(5) & 100    \\
1076   & $(9/2^-,11/2^+)$ & 335.2(5) & 3.6(7)  \\
       &           & 1049.3(5)& 9.1(9)  \\
       &           & 1076.3(5)& 19.5(13)\\
1425   & $(13/2^-)$& 348.7(5) & 42.8(14)\\ 
       &           & 684.7(5) & 84.5(23)\\
       &           & 768.8(5) & 29.5(15)\\
1473   & $(17/2^-)$& (48)$^b$& 7.5$^b$\\
       &           & 732.1(5)& 15.4(11)\\
X      &           &          &          \\
X+593  &           & 593.3(5) & 6.5(8)  \\
X+976  &           & 382.4(5) & 5.5(7)  \\
X+1365 &           & 389.5(5) & 8.8(9)  \\
\hline
\end{tabular}
\end{center} 
$^a$Transition not placed in the level scheme.\\
$^b$Transition not observed experimentally but deduced from the intensity balance
at the 1425-keV level. The relative intensity is calculated assuming an $E2$ 
multipolarity for the 48-keV transition and $E1$ multipolarity for the de-exciting
transitions.
\end{table}
In the present work, the weak 732-keV transition is observed in anti-coincidence 
with the 685-keV transition, which suggests an ordering of the 685-keV and 
713-keV transitions opposite to that suggested in \cite{St09}. Also, no 
experimental evidence was found for the 714-717-keV doublet reported in 
Ref.~\cite{St09}. A level scheme based on the coincidence studies performed in 
the present work, is shown in Fig.~\ref{123ag}. Table~\ref{123Ag} lists 
the $\gamma$-ray energies and intensities, as observed in the present study. 

\begin{figure}[t]
\begin{center}
\rotatebox{0}{\scalebox{0.3}[0.3]{\includegraphics{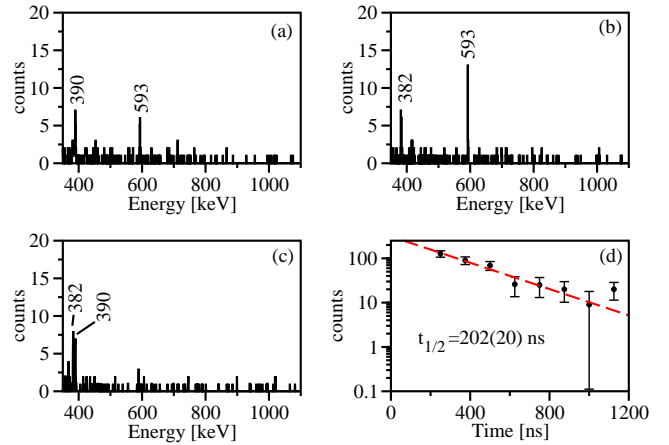}}} 
\caption[]{(color on-line) $^{123}$Ag $\gamma$-rays, observed in coincidence 
with the (a) 382-keV, (b) 390-keV, and (c) 593-keV transition, respectively.  
(d) a summed time spectrum for the three transitions.}
\label{coinc382_389_593}
\end{center}
\end{figure}

The delayed coincidence method \cite{An82} was used to estimate the half-lives of 
the 656-keV and 740-keV levels. It shows that the half-life of the two excited 
states is shorter than the DGF time binning, which is 25 ns per channel. Such a 
short half-life is consistent with a dipole or quadrupole nature for the 
629-keV and 713-keV transitions.

Coincidence spectra for the weak 382-keV, 390-keV and 593-keV transitions are 
shown in Fig.~\ref{coinc382_389_593}. The three $\gamma$-rays are in mutual 
coincidence. A half-life of $t_{1/2}=202$(20) ns is obtained from the fit to the 
summed time distributions for the 382-keV, 390-keV and 593-keV transitions. 
In the level scheme, this sequence is placed in parallel to the $\gamma$-rays 
de-exciting the 393-ns isomer, given that no experimental evidence was found for 
coincidences between the two branches. 

Two weak transitions with energies of 741 keV and 1008 keV are also observed in 
coincidence with the $^{123}$Ag ions, as shown in Fig.~\ref{123agene}. Due to 
the poor statistics no evidence for coincidences with the strongest gamma 
transitions was found. However, the two unplaced transitions may be related to a 
weak and fragmented $\gamma$-decay branch de-exciting the 1076-keV level. Such a 
scenario is supported by the observed intensity dis balance for the 1076-keV 
level.

The spin and parity assignments to the states in $^{123}$Ag are based on the 
$\gamma$-ray decay pattern, observed in the present work, and on the systematics.
\begin{figure}[t]
\scalebox{0.32}[0.32]{\includegraphics{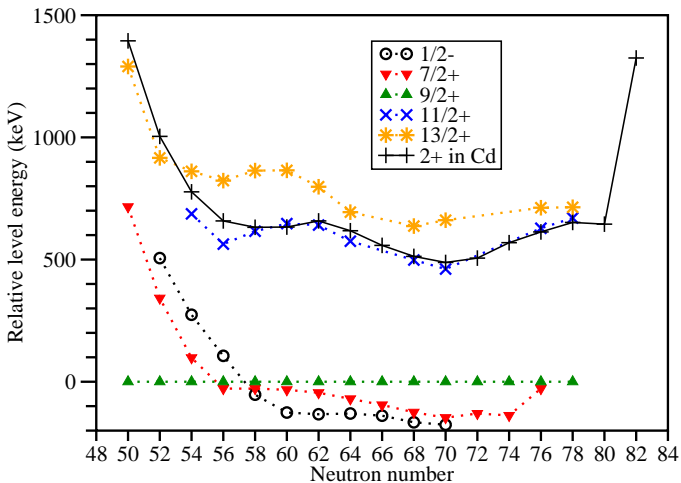}}
\caption[]{\label{e-n}(color on-line) Systematics of the low-lying states in the 
odd-mass Ag nuclei. Level energies are plotted relative to the $9/2^+$ level energy.}
\end{figure}

$\underline {7/2^+, 9/2^+}$: The systematics in Fig.~\ref{e-n} show that 
$9/2^+$ is the ground state in the $N=50-54$ silver nuclei, while 
$7/2^+$ is the first excited state and $1/2^-$ is the second excited state. 
$7/2^+$ becomes the ground state in $^{103}_{\ 47}$Ag$_{56}$ \cite{Fr09} and is 
the lowest lying positive parity state until $^{121}_{\ 47}$Ag$_{73}$ \cite{St09}. 
The ground state for the $N=58-70$  silver nuclei is $1/2^-$. The 
$1/2^-$ level energy in the heavier silver isotopes is unknown. Thus, based on 
the systematics, $1/2^-$, $7/2^+$ and $9/2^+$ are the most probable candidates 
for the ground state in $^{123}$Ag. However, because of the strong $\beta$-decay 
feeding to the 264-keV $(5/2^+,7/2^+)$ level in $^{123}$Cd, the $1/2^-$ and 
$9/2^+$ assignments to the ground state have been ruled out \cite{Oh04}. 
Given also that the fission process populates mainly yrast states, $(7/2^+)$ is 
assigned to the lowest lying state of the strongly populated sequence, 
de-exciting the 393-ns isomer. Again, based on the systematics, $(9/2^+)$ 
is adopted for the 27-keV level. 

$\underline {11/2^+, 13/2^+}$: 
Fig.~\ref{e-n} shows that the 11/2$^+$ state appears at approximately 600 keV 
above the $9/2^+$ state and that it is correlated to the 
$2^+$ level energies in the even-even cadmium nuclei. Also, in all silver 
isotopes with more than two valence neutrons outside the $N=50$ shell closure 
the 11/2$^+$ state is lower in energy than the $13/2^+$ state. Therefore, based 
on the systematics, $(11/2^+)$ and $(13/2^+)$ were assigned to the 656-keV and 
740-keV levels, respectively. These assignments are also consistent with the 
dipole or quadrupole nature of the 629-keV and 713-keV transitions, discussed above.

\begin{figure}[t]
\begin{center}
\rotatebox{0}{\scalebox{0.3}[0.3]{\includegraphics{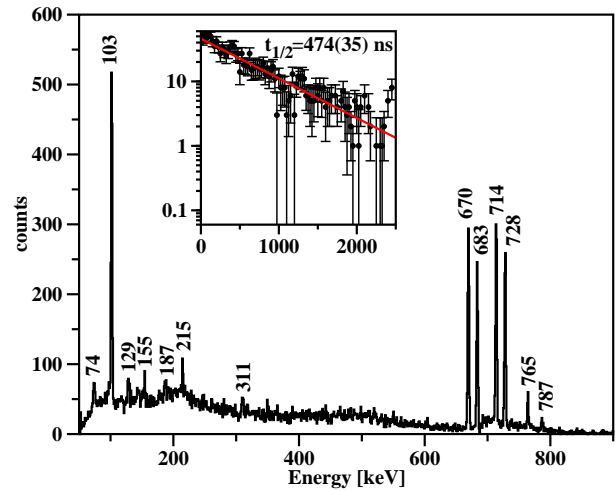}}} 
\caption[]{(color on-line) Gamma-ray energy spectrum, observed in coincidence 
with the $^{125}$Ag ions. (inset) A summed time spectrum for the 103-keV, 
683-keV and 728-keV transitions.}
\label{125agene}
\end{center}
\end{figure}
$\underline {13/2^-, 17/2^-}$: The 1473-keV level decays via a weak 732-keV
 transition to the ($13/2^+$) state, but not to the levels with lower spin. 
Therefore, $J^\pi\geq 17/2^\pm$ can be expected for the 1473-keV state. 
Similarly, the 1425-keV level does not decay directly to the $(9/2^+)$ level. 
Therefore, a tentative $J^\pi\geq 13/2^\pm$ assignment can be made for this 
level. Furthermore, in order for the 1473-keV level to be an isomeric state, a 
low-energy transition is expected to compete with the 732-keV transition. Such a 
transition would be a 48-keV transition linking the 1473-keV level with the 
1425-keV state. For a 48-keV transition of $E1$ or $E2$ nature the conversion 
coefficients calculated with BrIcc \cite{Bricc} are $\alpha (48\gamma; E1) =1.268$ and 
$\alpha (48\gamma; E2)=19.9$ and the hindrance factors would be 
$B(E1)=1.45\times 10^{-6}$ and $B(E2)=6.8 (8)$ W.u., respectively. Both values 
are consistent with the hindrance factor systematics in ref.~\cite{La11}, but 
for higher multipolarities the 48-keV transition would be enhanced from three to 
nine orders of magnitude and hence very unlikely. Also, the intensity balance, 
performed for the 1425-keV level, suggests that the 48-keV $\gamma$-ray intensity 
will be 69\% of the intensity of the 713-keV transition in case of $E1$ 
multipolarity, and 7.5\% if it is an $E2$ transition. Because, the 48-keV 
transition is not observed in the present experiment, $E2$ multipolarity was 
adopted. Hence, $J^\pi=(17/2^-)$ and $J^\pi=(13/2^-)$ assignments are made to 
the 1473-keV and 1425-keV states, respectively. Thus, assuming pure $E2$ and
$M2$ nature of the isomeric transitions in $^{123}$Ag, 
$B(E2; 17/2^-\rightarrow 13/2^-)=6.8 (8)$ W.u. and 
$B(M2; 17/2^-\rightarrow 13/2^+)=1.36(12)\times 10^{-3}$ W.u. 
A similar isomeric decay has been observed in $^{123}$In  with 
$B(E2;17/2^-\rightarrow 13/2^-)$=3.3(5) W.u. \cite{Sc04}.

$\underline {9/2^-,11/2^+}$: Given that the 1076-keV level decays to the 
$(7/2^+)$ ground state and is fed by the 349-keV transition from the $(13/2^-)$ 
state, a $(9/2^-,11/2^+)$ assignment to the level was made.

\subsubsection{$^{125}$Ag}
\begin{figure}[t]
\begin{center}
\rotatebox{0}{\scalebox{0.35}[0.35]{\includegraphics{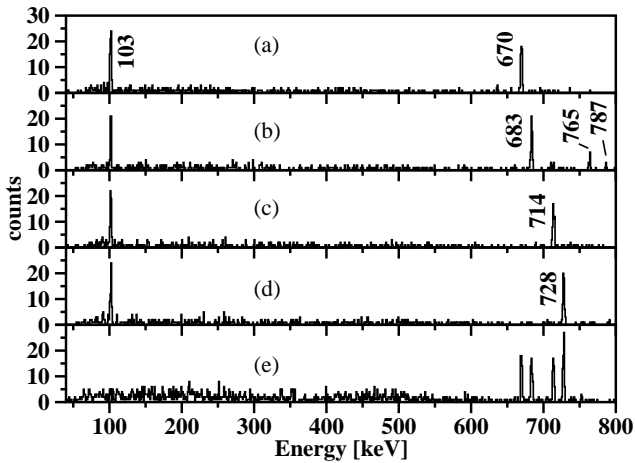}}} 
\caption[]{$^{125}$Ag $\gamma$-rays, observed in coincidence with the (a) 
728-keV, (b) 714-keV, (c) 683-keV, (d) 670-keV, and (e) 103-keV transitions, 
respectively.}
\label{125agcoinc}
\end{center}
\end{figure}

In the present work, $^{125}$Ag was observed in the $^{120}$Rh FRS setting from 
$^{238}$U fission and in the $^{130}$Cd FRS setting from the fragmentation of 
the $^{136}$Xe beam. Fig.~\ref{125agene} shows $\gamma$-rays observed in 
coincidence with the $^{125}$Ag ions within a 3.85 $\mu$s wide time window 
opened 125 ns after implantation. A group of four intense transitions at energies 
of approximately 700 keV and a strong 103-keV $\gamma$-line is observed. 
Coincidence spectra are shown in Fig.~\ref{125agcoinc}. Fig.~\ref{125agcoinc}(a) 
presents coincidences between the 103-keV and the 728-keV transitions.

 The 765-keV 
$\gamma$-rays are only in coincidence with the 714-keV $\gamma$-rays, as shown 
in Fig.~\ref{125agcoinc}(b). 
\begin{figure}[t]
\begin{center}
\rotatebox{90}{\scalebox{0.3}[0.3]{\includegraphics{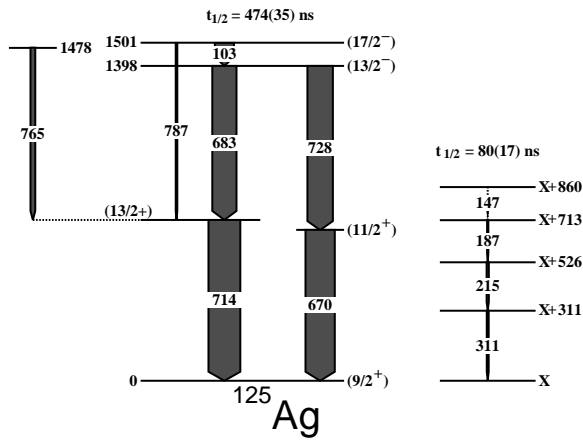}}} 
\caption[]{Partial level scheme of $^{125}$Ag, based on the observed $\gamma$-ray coincidences.}
\label{125ag}
\end{center}
\end{figure}
Fig.~\ref{125agcoinc}(c) shows coincidences between the 683-keV and 714-keV 
transitions and Fig.~\ref{125agcoinc}(d) -- the coincidences between 670-keV and
728-keV transitions. Fig.~\ref{125agcoinc}(e) shows that the 103-keV 
$\gamma$-rays are in coincidence with the most intense $\gamma$-rays but not in 
coincidence with the 765-keV and 787-keV transitions. 

The level scheme, presented in Fig.~\ref{125ag}, is based on the coincidences 
observed in the present work. The energy difference between the 1501-keV and 
1478-keV levels is 23 keV, and hence impossible to detect with the set up used.
Also, because of the poor statistics, no half-life information was obtained from 
the time distribution of the 765-keV transition, which makes it 
difficult to determine whether the 1478-keV level is fed by the 1501-keV isomer 
or it is a different isomeric state. 

\begin{figure}[t]
\begin{center}
\rotatebox{0}{\scalebox{0.31}[0.31]{\includegraphics{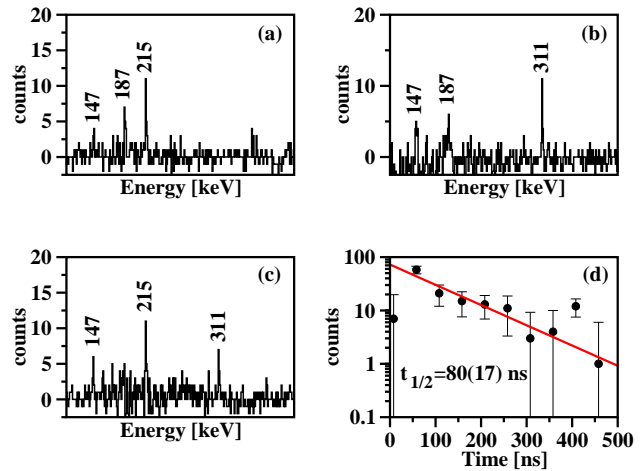}}} 
\caption[]{(color on-line) $^{125}$Ag $\gamma$-rays, observed in coincidence 
with the (a) 311-keV, (b) 215-keV and (c) 187-keV transition, respectively. (d)
a summed time spectrum for the three transitions.}
\label{t187_215_311}
\end{center}
\end{figure}

The four transitions with energies of 670 keV, 683 keV, 714 keV and 728 keV were 
previously observed and placed in a level scheme \cite{St09}. Based on the 
coincidences between the 787-keV and 714-keV $\gamma$-rays, the ordering of the 
714-keV and 683-keV sequence is inverted with respect to \cite{St09}. 

The inset of Fig.~\ref{125agene} shows a summed time spectrum for the isomeric 
state in $^{125}$Ag and the fit to the slope of the distribution  
gives a half-life of 474(35) ns for the 1501-keV isomeric state. The half-life, 
obtained in the present work, overlaps with $t_{1/2} = $473(111) ns given in 
\cite{St09} and with 0.44(9) $\mu$s from \cite{To06} and is consistent with
$0.498^{+21}_{-20}$ $\mu$s in \cite{Ka12}.

The delayed coincidence method \cite{An82} was used to estimate the half-life of 
the 714-keV and 670-keV levels. It showed that the half-life of the two 
excited states is shorter than the DGF time binning. This observation is 
consistent with a dipole or a quadrupole nature of the 714-keV and 670-keV 
transitions.

Several weak lines with energies of 74 keV, 129 keV, 155 keV, 187 keV, 215 keV, 
and 311 keV were observed in coincidence with the $^{125}$Ag ions. The 311-keV, 
215-keV and 187-keV lines were observed in mutual coincidence and in a 
tentative coincidence with a weaker 147-keV transition as shown in 
Fig.~\ref{t187_215_311}(a), (b) and (c). The half-life of 80(17) ns 
was deduced from the slope of the time spectrum shown in 
Fig.~\ref{t187_215_311}(d). Coincidences between this group of transitions and
the transitions de-exciting the 474-ns isomer were not observed, which enables
the placement of this group of transitions in parallel to the main branch.

\begin{table}[t]   
\begin{center}
\caption{\label{125Aglist} Level energies ($E_i$), spin/parity ($J^\pi$) 
assignments, $\gamma$-ray energies ($E_\gamma$) and relative intensities 
($I_{rel}$) for transitions, observed in coincidence with the 
$^{125}$Ag ions.}
\begin{tabular}{ccccc} 
\hline\hline
$E_i$ (keV) & $J^\pi$ &   $E_\gamma$ (keV) &$I_{rel}$ (\%)\\
\hline
      &            &    74.2(5)$^\dagger$ & 8(1)    \\
      &            &  129.3(10)$^\dagger$& $\leq$ 1\\
      &            &  155.0(10)$^\dagger$& $\leq$ 1\\
0     & $(9/2^+)$  &            &         \\
670   & $(11/2^+)$ &  669.8(5)  & 90(3)   \\
714   & $(13/2^+)$ &  714.1(5)  & 100     \\
1398  & $(13/2^-)$ &  683.4(5)  & 73(3)   \\
      &            &   728.3(5)  & 80(3)   \\
1478  &            &  764.5(5)  & 13.8(16)\\
1501  & $(17/2^-)$ & 102.5(5)  & 61.6(20)\\
      &            & 787.0(5)  & 5.5(12) \\
 X    &            &            &         \\
X+311 &            &  310.6(5)  & 6.0(12) \\
X+526 &            &  215.0(5)  & 8(4)    \\
X+713 &            &  187.3(5)  & 3.9(13) \\
X+860 &            &  147.0(10) & $\leq$ 1\\
\hline
\end{tabular}
\end{center} 
$^\dagger$Not placed in the level scheme. 
\end{table}
The $\gamma$-ray energies and intensities are listed in Table~\ref{125Aglist}. 
An intensity imbalance between the 714-keV and 683-keV transitions was reported 
in \cite{St09}, with the 683-keV transition being the stronger. The 
analysis, performed in the present work, shows indeed an imbalance of 17\%, but 
in the opposite direction. However, the 765-keV and 787-keV transitions were 
also observed in coincidence with the 714-keV transition, helping to balance the 
intensity for the 714-keV level.
The intensity balance for the 1398-keV level suggests that the 103-keV 
transition is converted with $\alpha = 1.48(8)$, which is close to the theoretical
value of 1.3 for an $E2$ transition calculated with BrIcc \cite{Bricc}. 

In $^{125}$Ag, the spin and parity assignments are based on the observed  
$\gamma$-decay pattern, on analogy with $^{123}$Ag, and on the systematics 
presented in Fig.~\ref{e-n}. In contrast to the level scheme proposed in 
\cite{St09, Ka12}, $J^\pi =11/2^+$ was assigned to the lower-lying excited state 
in $^{125}$Ag. For the 1501-keV level, the 
$B(E2; 17/2^-\rightarrow 13/2^-)$=1.18(11) W.u. and 
$B(M2; 17/2^- \rightarrow 13/2^+)=3.2(8) \times 10^{-4}$ W.u. values, calculated 
with $\alpha (103\gamma ;E2)=1.324$  and $\alpha (787\gamma ;M2)=0.0056$ 
\cite{Bricc}, are consistent with the hindrance factor systematics \cite{La11}.

\subsection{Even-A Ag nuclei}

\begin{figure}[t]
\begin{center}
\rotatebox{90}{\scalebox{0.3}[0.3]{\includegraphics{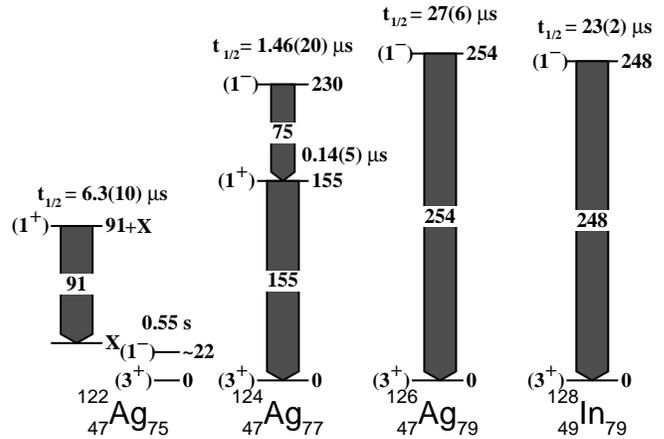}}} 
\caption[]{$^{122,124,126}$Ag level schemes compared to the $^{128}$In level 
scheme reported in \cite{Sc04}. The $(1^-)$ level in $^{122}$Ag is from 
\cite{Ta07}.}
\label{levels122ag}
\end{center}
\end{figure}

Fig.~\ref{levels122ag} presents the partial level schemes of $^{122}$Ag,
$^{124}$Ag and $^{126}$Ag, obtained in the present work and compared to the
odd-odd neighbor $^{128}$In \cite{Sc04}.

\subsubsection{$^{122}$Ag}
Prior to our study two isomeric states with half-lives of 0.20 s and 
0.55 s and $J^\pi=(9-)$ and $(1-)$ were observed in $^{122}$Ag \cite{Kr00, Ta07},
but no isomeric decay transitions had been seen. The present work reports on a 
new shorter-lived isomer, observed in the relativistic fission of $^{238}$U and 
the fragmentation of $^{136}$Xe. In these experiments the FRS was tuned to 
transmit $^{120}$Rh and $^{126}$Cd respectively.

\begin{figure}[t]
\begin{center}
\rotatebox{0}{\scalebox{0.3}[0.3]{\includegraphics{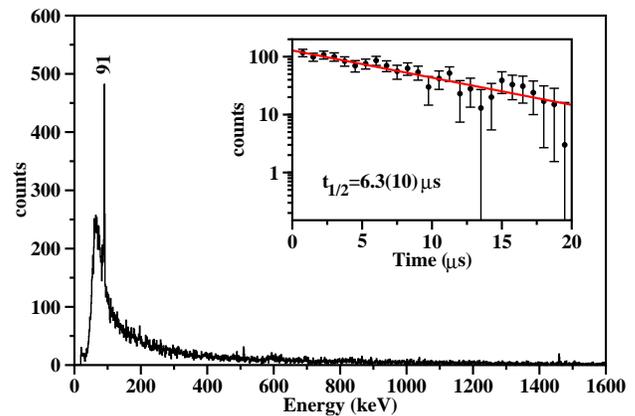}}} 
\caption[]{(color on-line) Gamma-ray energy spectrum observed in coincidence with $^{122}$Ag ions. 
The inset shows the time distribution of the 91-keV transition.}
\label{enetime122}
\end{center}
\end{figure}

Fig.~\ref{enetime122} shows a  $\gamma$-ray spectrum observed in delayed 
coincidence with the $^{122}$Ag ions within a 8.75-$\mu$s wide time window 
opened 2.75 $\mu$s after implantation. A single $\gamma$-ray with an energy of 
91 keV is observed.  The inset of Fig.~\ref{enetime122} shows the time distribution 
of the 91-keV transition. A half-life of $t_{1/2}=6.3$(10) $\mu$s was obtained 
from the slope, which suggests $E1$ or $E2$ multipolarity, given that only
$B(E1)=4.8(8)\times 10^{-8}$ W.u. and $B(E2)=0.133(23)$ W.u. calculated with 
$\alpha (91\gamma; E1)=0.2092$ and $\alpha (91\gamma; E2)=2.02$ \cite{Bricc} are 
consistent with the hindrance factor systematics \cite{La11}. 

Similar low-lying $\mu$s isomeric states were observed in the 
neutron-rich $^{126,128,130}$In \cite{Sc04} and related to the decay of the 
$1^-$ and $3^+$ states. Also, an excited $1^+$ state is present in 
$^{126,128,130}$In \cite{Sc04}. It is 1692 keV above the $3^+$ level in 
$^{130}$In and decreases in energy to 688-keV above the $3^+$ ground state in 
$^{126}$In. In $^{122}$Ag, however, $3^+$ is assigned to the ground 
state \cite{Ta07} on the basis of the log{\it ft}$\approx 5.4$ and 5.9 to the 
daughter 2+ and (4+) levels, respectively, and $(1^-)$ is assigned to the 
0.55-s isomer in \cite{Kr00}. Therefore, based on the analogy with 
$^{126,128,128}$In, $J^\pi =(1^+)$ was tentatively assigned to the isomeric 
level observed in the present study. Given that the multipolarity of the 91 keV 
transition is $E1$ or $E2$, the final state can be the $(1^-)$ excited state or 
the $(3^+)$ ground state.

The level energy $E(1^-)\approx 22$ keV of the 0.55-s isomer is estimated from 
its half-life and the $B(M2)$ transition strength in $^{126,128}$In \cite{Sc04}. 
This enables the placement of the $(1^-)$ state in between the $(1^+)$ isomeric 
state and the ground state of $^{122}$Ag. 

\subsubsection{$^{124}$Ag}
Prior to our study, two $\gamma$-rays with energies of 155 keV and 1132 keV were
observed in mutual coincidence and associated with $^{124}$Ag \cite{To06}. 
\begin{figure}[t]
\begin{center}
\rotatebox{0}{\scalebox{0.35}[0.35]{\includegraphics{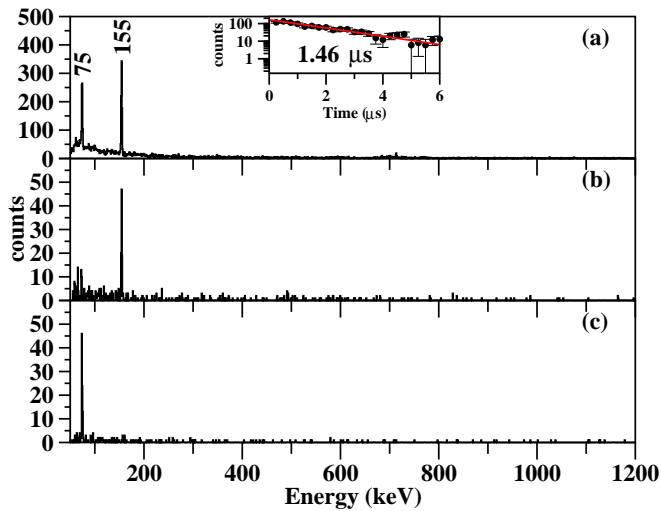}}} 
\caption[]{(color on-line) Gamma-ray energy spectra observed in coincidence with $^{124}$Ag 
ions. (a) total projection; (b) $\gamma$-rays in coincidence with the 75-keV transition;  
(c) $\gamma$-rays in coincidence with the 155-keV transition; (inset) a summed time spectrum 
for the 75-keV and the 155-keV transitions.}
\label{ene124}
\end{center}
\end{figure}

In the present work, $^{124}$Ag was observed in the $^{120}$Rh FRS settings 
from $^{238}$U fission and in the $^{126}$Cd and $^{130}$Cd FRS settings from 
fragmentation of the $^{136}$Xe beam. Fig.~\ref{ene124} shows $\gamma$-ray 
spectra, observed in delayed coincidence with $^{124}$Ag nuclei within a 5.75 
$\mu$s time window opened 450 ns after implantation. Two transitions with energies 
75 keV and 155 keV were observed to be in coincidence. The two $\gamma$-rays are
observed also in \cite{Ka12}. The 1132-keV line, reported previously in 
\cite{To06}, is not confirmed by the present study. The inset of 
Fig.~\ref{ene124}(a) shows the summed time spectrum for the 75-keV and 155-keV 
transitions and the half-life, obtained from the fit to the slope is 
1.46(20) $\mu$s, which is consistent with the 1.62$^{+29}_{-24}$ $\mu$s in \cite{Ka12}. 

\begin{figure}[t]
\scalebox{0.66}[0.66]{\includegraphics{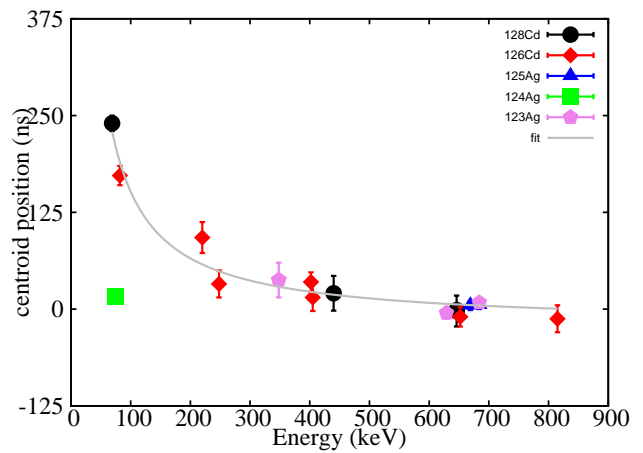}}
\caption[]{\label{cs}(color on-line) Time-walk correction for the Ge detectors.
Details about the $^{124}$Ag data point are presented in the text.}
\end{figure}

The 155-keV $\gamma$-ray is delayed with respect to the 75-keV $\gamma$-ray, 
which enables the ordering made in the level scheme in Fig.~\ref{levels122ag}.
However, the two transitions are of low energy and therefore a time-walk 
correction has to be made in order to properly determine the half-life of the 
intermediate state. The time-walk of the prompt distribution as a function 
of the energy was determined by using data for $^{126,128}$Cd from 
\cite{Ho07, Ca09} and for $^{123,125}$Ag from the present experiment. Time 
distributions, calculated as the difference $T=t_1-t_2$ between the detection 
time of the feeding $t_1$ and de-exciting $t_2$ transitions, were analyzed. The 
centroids of the prompt distribution were obtained for several transitions 
in $^{126,128}$Cd and $^{123,125}$Ag where the level of interest is fed by a 
low-energy transition and de-excited by a high-energy transition. The 
centroid of the time distribution is plotted on Fig.~\ref{cs} as a function of 
the low-energy feeding transition. The energy-time dependence was fitted with a 
function of the type $T(E_\gamma)=a/E_\gamma+b$, where $T$ is the centroid of 
the time distribution, $E_\gamma$ is the energy of the feeding transition in keV 
and $a=17228$ ns$\cdot$keV, $b=-20.6$ ns are the parameters obtained from the 
fit. A mirror symmetric function \cite{JM10} for the position of the prompt 
distribution can be obtained from a set of states fed by a high-energy 
transition and decaying via a low-energy transition. In the $^{124}$Ag case, the 
two transitions have low energies. Therefore, the $^{124}$Ag data point was 
plotted on Fig.~\ref{cs} after time-walk correction was applied for the 
75-keV feeding transition. Thus, a half-life of 0.14(5) $\mu$s was obtained 
from the difference between the $^{124}$Ag data point and the $T(E_\gamma)$ value
for $E_\gamma$=155 keV. The $B(E1)=5.0(18)\times 10^{-7}$ W.u. and $B(E2)=0.9 (4)$ W.u.,
values, calculated for the 155-keV transition with $\alpha (155\gamma;E1)=0.0457$ 
and $\alpha (155\gamma;E2)=0.303$ \cite{Bricc}, are consistent with the systematics 
\cite{La11}. An $M1/E2$ multipolarity assignment for the 155-keV transition is 
also possible for $\delta \le 0.1$.

Table~\ref{124Agenint} shows that the 75-keV transition has an intensity of 
74(38)\% of that of the 155-keV transition. The intensity balance for the 
155-keV level leads to $\alpha (75\gamma)=0.4(8)$ in case of pure $E1$ 155-keV
transition and to $\alpha (75\gamma)=0.8(10)$ if 155-keV transition is a pure 
$E2$ transition. These conversion coefficients are consistent with an $E1$ 
nature of the 75-keV transition. Thus, $B(E1)=3.2(5)\times 10^{-7}$ W.u. was 
obtained with $\alpha (75\gamma;E1)=0.363$. 

\begin{table}[t]  
\caption{\label{124Agenint}Decay properties of the isomeric state in $^{124}$Ag. 
Level energy ($E_i$), half-life 
($T_{1/2}$), $\gamma$-ray energy $(E_\gamma)$, relative intensity ($I_\gamma$) and 
the experimental total conversion $\gamma$-ray coefficient ($\alpha_{exp}$) 
obtained for the 75-keV transition and compared to the theoretical total 
conversion coefficients ($\alpha _{th}$) calculated for $E1$, $M1$, $E2$ and 
$M2$ multipolarities.}
\begin{center}
\begin{tabular}{ccccccccccc} 
\hline\hline
$E_i$ & $T_{1/2}$ & $E_\gamma$   & $I_\gamma$ & $\alpha _{exp}$ & $\alpha _{th}$ & $\alpha _{th}$ & $\alpha _{th}$ & $\alpha _{th}$\\
(keV)& ($\mu$s)  & (keV) &\%  &                 & $E1$ & $M1$  & $E2$ & $M2$\\
\hline
230 & 1.46 & 75    & 74(38) & 0.4(8);0.8(9) & 0.36 & 0.95 & 4.0  & 13\\
155 & 0.14 & 155   &100(16) &               & 0.05 & 0.12 & 0.30 & 0.89\\
\hline\hline
\end{tabular}
\end{center} 
\end{table}

\begin{figure}[th]
\begin{center}
\rotatebox{0}{\scalebox{0.3}[0.3]{\includegraphics{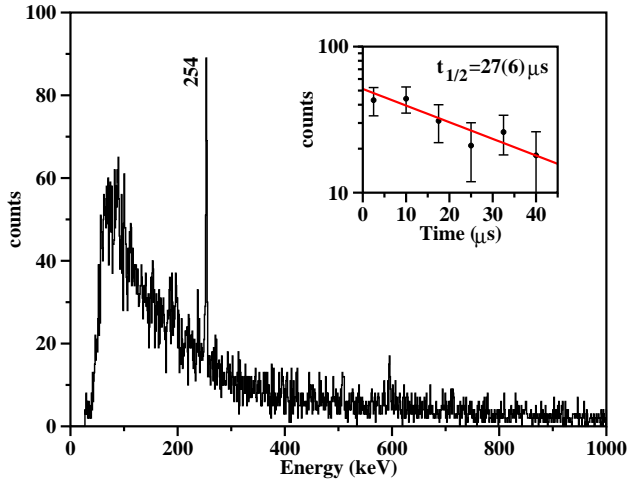}}} 
\caption[]{\label{ene126}(color on-line) Gamma-ray energy spectrum observed in 
delayed coincidence with the $^{126}$Ag ions. (inset) Time spectrum gated on the 
254-keV transition.}
\end{center}
\end{figure}

Since $J^\pi \ge 2$ was assigned to the ground state of $^{124}$Ag \cite{KW08} and 
given no other experimental data is available, the spin/parity assignments to the 
levels in $^{124}$Ag are based on analogy with $^{122}$Ag and the systematics 
for the neutron-rich indium nuclei \cite{Sc04}.

\subsubsection{$^{126}$Ag}
$^{126}$Ag is the most neutron-rich nucleus in the silver isotopic chain studied 
in the present experiment. The ions were transmitted through the FRS in the 
$^{130}$Cd setting during the $^{136}$Xe beam fragmentation experiment. The 
energy spectrum, shown on Fig.~\ref{ene126}, was incremented for delayed 
transitions in coincidence with the $^{126}$Ag ions within a 48-$\mu$s wide 
time window placed 0.58 $\mu$s after the prompt-$\gamma$ flash. A single 
$\gamma$ ray at energy of 254 keV was detected by RISING. The inset of the 
figure presents the time distribution of the 254-keV line and a half-life of 
27(6) $\mu$s was obtained from the slope. The isomeric transition in $^{126}$Ag 
is also reported in \cite{Ka12} and a lower limit of the half-life is given to 
be 20 $\mu$s.

An $M2$ multipolarity assignment for the isomeric transition is based on 
comparison between the measured half-life with the Weisskopf estimates for the 
partial half-life of a 254-keV transition, and the hindrance factor systematics 
for low-lying transitions in the mass region $100<A<132$ \cite{La11}. 
$B(M2)=0.037(9)$ W.u. is calculated with $\alpha (254\gamma;M2)=0.1645$ \cite{Bricc}. 
The $M2$ nature of the isomeric transition is also consistent with the isomeric 
decay in the $^{128}_{\ 49}$In$_{79}$ isotone as shown in Fig.~\ref{levels122ag}, 
where the $J^\pi=(1^-)$, $t_{1/2}$=23 $\mu$s isomer decays via a 248-keV $M2$ 
transition to the $(3^+)$ ground state with $B(M2)=0.047(5)$ W.u. Therefore,
$(1^{-})$ and $(3^+)$ are assigned to the 254-keV and the ground state levels in 
$^{126}$Ag.

\section{Discussion}

\subsection{odd-$A$}

The systematics, presented in Fig.~\ref{e-n}, show that the level energies in 
the odd-mass silver nuclei evolve smoothly with the neutron number following the
trend of the first $2^+$ level of the Cd core. Often, such states are interpreted 
as core coupled states. 

Based on the relative positioning of the $7/2^+$ level with respect to the 
$9/2^+$ level, two groups of nuclei can be distinguished. For $N\le 54$ the 
$7/2^+$ level energy is higher than the $9/2^+$ level energy. For $N>54$ the 
relative positioning of the two levels inverts. Extrapolating the trend 
towards $N=82$, it can be expected that $9/2^+$ becomes the ground state again. 
While the group of nuclei with less then six valence particles or holes can be 
understood in the single-particle framework the anomalous position of the 
$j-1=7/2$ state hints at a more complex structure.
\begin{figure}[t]
\scalebox{0.32}[0.32]{\includegraphics{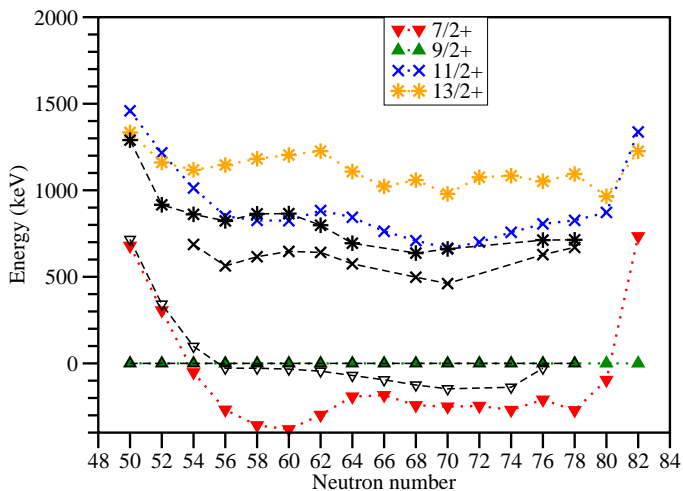}}
\caption[]{\label{esmal}(color on-line) Evolution of the $\pi g_{9/2}^{-3}$ 
cluster in odd-A Ag nuclei. The theoretical level energies are plotted with 
colored symbols and compared to the experimental level energies in black. 
The same symbols are used for theoretical and experimental levels of a given 
spin and parity.}
\end{figure}
\begin{figure*}[t]
\begin{center}
\rotatebox{90}{\scalebox{0.6}[0.6]{\includegraphics{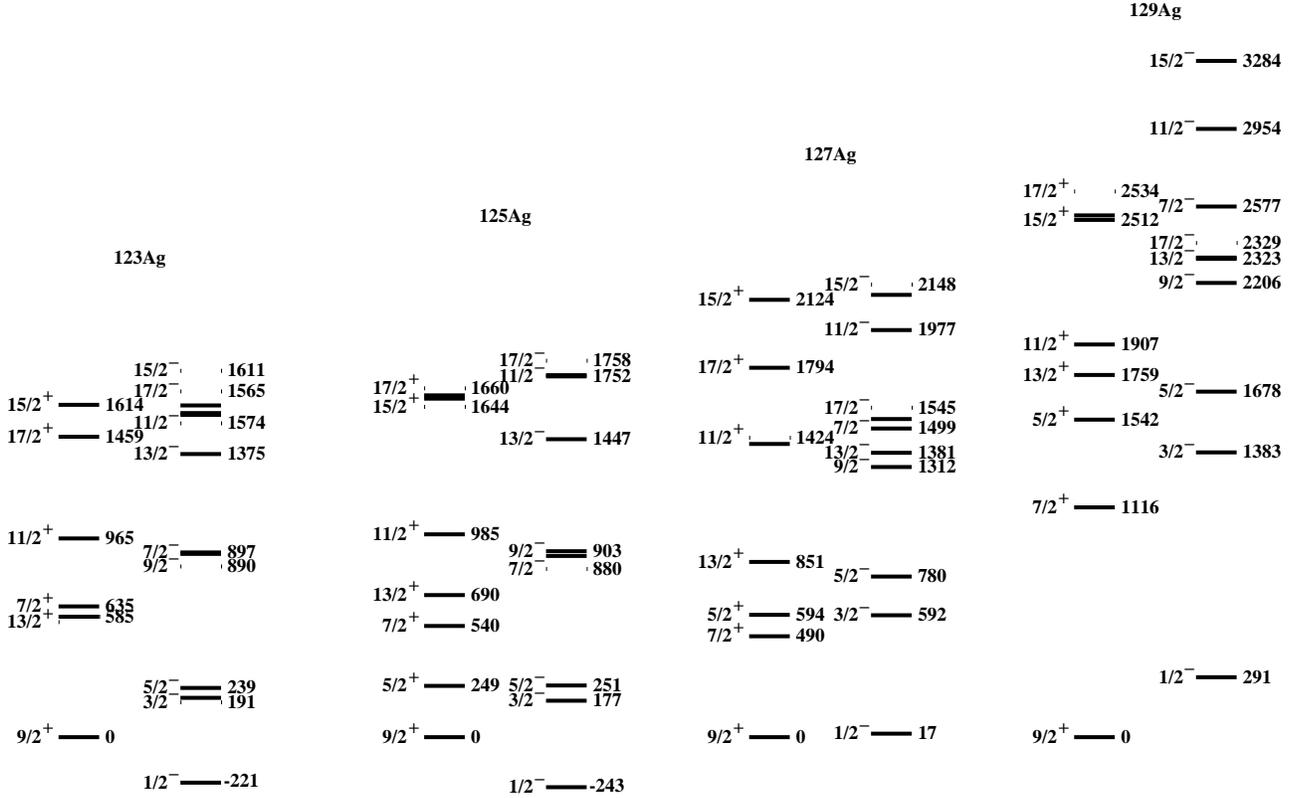}}} 
\caption[]{\label{NuShell}The results of Shell-model calculations for 
$^{123, 125, 127, 129}$Ag, obtained with the jj45pna interaction.}
\end{center}
\end{figure*}

The origin of this anomaly was initially related to the three-proton cluster 
configuration \cite{Ki66}. Having 47 protons, the only available proton sub-shell 
for the silver valence proton holes is $1g_{9/2}$. Also, given that this is a
unique-parity orbit, the low-lying positive parity states can arise only from 
the $\pi g_{9/2}^{-3}$ configuration. By using the expansion coefficients of the 
three-particle matrix elements of a two-body interaction in terms of 
two-particle matrix elements for the $\{j^{-3}\}$ 
configuration of identical nucleons \cite{He94}, energy spectra were calculated 
for the $^{97-129}_{\ \ \ \ \ 47}$Ag$_{50-82}$ isotopes. Fig.~\ref{esmal} shows 
the energy spectra, obtained from
\begin{equation}
<j^3\alpha ;JM|H|j^3\alpha ;JM> = 3\sum_{J'}[j^2(J')jJ|\}j^3J]^2A_{J'},
\end{equation}
where $A_{J'}$ are the two-body matrix elements, parametrized with respect to the
neighboring even-even cadmium nuclei, and $[j^2(J')jJ|\}j^3J]$ are the 
coefficients of fractional parentage listed in \cite{BK71}. With the exception of  
$^{101}$Ag$_{54}$ and $^{125}$Ag$_{78}$, this simple approach gives a good 
overall description of the experimental data and in particular it reproduces 
correctly the sequence of the levels, hinting at the importance of the 
$\pi g_{9/2}^{-3}$ cluster configuration in understanding the structure of the
silver nuclei. However, the systematics in Fig.~\ref{esmal} show also that the 
theoretical $11/2^+$ and $13/2^+$ level energies are systematically 
overestimated and the description of the $7/2^+$ level energy in 
$^{103-109}_{\ \ \ \ \ \ 47}$Ag$_{56-62}$ and in $^{123}_{\ 47}$Ag$_{76}$ is 
poor. Moreover, the negative-parity $13/2^-$ and $17/2^-$ levels in 
$^{123,125}$Ag, are outside the model space. A more realistic description of 
the excited states in the odd-mass silver isotopes, and in $^{123,125}$Ag 
nuclei in particular, should include a larger space and a certain degree of 
collectivity. A small quadrupole deformation was suggested already at 
$^{101}$Ag$_{54}$ to account for the $E3$ isomeric transition strength between 
the $1/2^-$ and $7/2^+$ states \cite{CKN75}. Further away from the $N=50$ magic 
number rotational bands based on the $\pi 7/2^+[413]$ Nilsson orbit appear 
\cite{Hw02}. As the $N=82$ magic number is approached, a reduction of 
collectivity should take place and the excited states are expected to have purer
wave functions. 

Given that the above approach does not take into account the proton-neutron and 
neutron-neutron interactions, which are important in the mid-shell and 
transitional regions, shell model calculations were performed for the 
$^{123-129}$Ag odd-even nuclei using the NuShell code \cite{Br04}. The full 
jj45pn space, involving $1f_{5/2}$, $2p_{3/2}$, $2p_{1/2}$, $1g_{9/2}$ proton and 
$1g_{7/2}$, $2d_{5/2}$, $2d_{3/2}$, $3s_{1/2}$, $1h_{11/2}$ neutron orbitals was 
used. The theoretical spectra, shown  in Fig.~\ref{NuShell} are obtained with the 
jj45pna interaction, parametrized with respect to the nuclei lying close to the 
doubly magic $^{132}$Sn. The theoretical level energies evolve smoothly from 
$^{129}$Ag to $^{123}$Ag, predicting the $9/2^+$ state to be the lowest-lying 
positive-parity state. In  $^{129}$Ag, the wave function of the $J^\pi=9/2^+$ 
ground state and the $J^\pi = 11/2^+ - 17/2^+$ yrast states consist of almost
 pure $\pi 1g_{9/2}^{-3}$ configuration with an amplitude of 68\% for the ground 
state. In the lighter $^{123,125}$Ag the respective wave functions are more 
fragmented, because of the larger valence space. Even though, the positive parity 
states observed in $^{123,125}$Ag are present in the calculated level schemes, 
their sequence is incorrect. In particular, the $7/2^+$ level energy in 
$^{123}$Ag appears 635 keV above the $9/2^+$ state and the ordering of the 
$11/2^+,13/2^+$ doublet is opposite to the experimental level scheme. Also, the 
energies of the isomeric $17/2^-$ and $13/2^-$ levels in $^{123,125}$Ag are 
overestimated by NuShell by 300 to 500 keV. Similar deviations from the 
experimental data were observed also in the neutron-rich indium isotopes 
\cite{Sc04}. The Shell model with the jj45pna interaction and  $e_p=1.35$ and 
$e_n=0.78$ effective charges gives an overestimated 
$B(E2;17/2^-\rightarrow 13/2^-)$=18.5 W.u. value for the isomeric 48-keV 
transition in $^{123}$Ag with respect to the experimentally observed 6.8(8) W.u. 
given in sect.III.A.1.

A different theoretical approach was given many 
years ago by a model where a cluster of three valence protons was coupled to 
quadrupole vibrations \cite{Pa73}. In these calculations, the particles are 
allowed to move in the $\pi g_{9/2}$, $\pi p_{1/2}$ and $\pi p_{3/2}$ sub-shells 
and are coupled to a Sn vibrational core. Even though, there are no specific 
calculations for the neutron-rich silver isotopes, a qualitative 
analysis can be made given that the even-even medium-mass cadmium nuclei have a 
smooth behavior with their first phonon being at approximately 500-600 keV. 
A good overall description of the medium-mass 
silver nuclei, where the $j-1$ anomaly takes place, can be obtained with a coupling 
parameter strength $a\ge 0.7$. The energy levels were calculated in \cite{Pa73} 
for the negative parity states up to $19/2_1^-$ and for the positive-parity 
states up to $13/2_1^+$. The $7/2_1^+$ level has a seniority $v=3$ zero-phonon 
and one-phonon component, as well as a $v=1$ one-phonon contribution, each with 
an amplitude of approximately 20\%. The $9/2_1^+$ level has a dominant 
contribution of the $v=1$ zero-phonon component with an amplitude of 30\% and a 
$v=3$ one-phonon component with a contribution of 17\%. The $11/2^+$ state is 
more fragmented with $v=3$ zero-phonon, $v=1$ one-phonon and $v=2$ two-phonon 
components with amplitudes of 16\%, 21\% and 12\%, respectively. For the 
$13/2^+$ state, the components with amplitudes higher than 10\% are the $v=3$ 
zero-phonon and the $v=1$ one-phonon. 

The model calculations give $B(E2;17/2_1^-\rightarrow 13/2_1^-)=18.4$ W.u. or 
16.0 W.u. values for the 48-keV isomeric transition in $^{123}$Ag, obtained with 
two sets of effective charges $e^{s.p.}_{eff}=2$ and $e^{vib}_{eff}=2$ and 2.5. 
Both values overestimate the experimental one. Nevertheless, the model gives a 
good qualitative description of the level sequence and the multiplet structure 
in particular, hinting at the importance of zero- and one- phonon excitations in 
understanding the structure of the lowest-lying positive-parity yrast states.

In the low spin regime, the cluster-vibration and the shell model predict 
different transition strengths. The shell model calculations give 
$B(E2,9/2_1^+\rightarrow 7/2_1^+)=4.8$ W.u. for the 27-keV transition in 
$^{123}$Ag, while the cluster-vibration model predicts a more 
enhanced transition with $B(E2,9/2_1^+\rightarrow 7/2_1^+)=27.5$ W.u. Even 
though, the present experiment does not allow the measurement of the half-life 
and the mixing ratio of the 27-keV transition in $^{123}$Ag, the models have 
clear predictions which can be tested in future experiments.

\subsection{even-A} 

The low-lying excited states in the odd-odd neutron-rich silver nuclei can be
described by using the cluster-vibration model, where the proton-neutron 
residual interaction is a result of quadrupole and spin vibration phonon exchange
between the odd particles and the nuclear core \cite{Pa79}. As a result, split 
multiplets with level energies $E[(j_p,j_n)J]$ as a function of the nuclear spin 
$J$ arise. The level energies obey the parabolic rule
\begin{equation}
E[(j_p,j_n)J]=E_{j_p}+E_{j_n}+\delta E_2 + \delta E_1 ,
\end{equation}
where the quadrupole $\delta E_2$ and spin-vibrational $\delta E_1$ 
contributions to the splitting of the multiplet are given by
\begin{widetext}
\begin{equation}
\begin{split}
\delta E_2 & = -\alpha _2\mathscr{V}{[J(J+1)-j_p(j_p+1)-j_n(j_n+1)]^2+J(J+1)-j_p(j_p+1)-j_n(j_n+1)
\over 2j_p(2j_p+2)2j_n(2j_n+2)}+\mathscr{V}{\alpha _2\over 12}\\
\delta E_1 & =-\alpha _1\xi {J(J+1)-j_p(j_p+1)-j_n(j_n+1)\over (2j_p+2)(2j_n+2)} \ .\\
\end{split}
\end{equation}
\end{widetext}
Here, $E_{j_p}$ and $E_{j_n}$ denote the quasi-proton $|j_p>$ and the quasi-neutron 
$|j_n>$ energies, and are deduced from the experimental data. $j_p$ and $j_n$ are 
the proton and neutron total angular momenta, and $J=|j_p-j_n|,...,(j_p+j_n)$ is 
the nuclear total angular momentum. If $|j_p>$ and $|j_n>$ are both particle-like 
or both hole-like the parameter $\mathscr{V}=1$, otherwise $\mathscr{V}=-1$. The 
parameter $\xi$ is defined as $\xi =1$ if the Nordheim number 
$\mathscr{N}=j_p-l_p+j_n-l_n=-1$ and 
$\xi = (2j_p+2)(2j_n+2)/(2j_p2j_n)$ if $\mathscr{N}=1$. Otherwise, 
$\xi =-(2j_p+2)/2j_p$ for $\mathscr{N}=0^-$ and $\xi = -(2j_n+2)/2j_n$ for 
$\mathscr{N}=0^+$, where the symbols $\mathscr{N}=0^-$ and $\mathscr{N}=0^+$ 
distinguish the cases $j_n-l_n=-1/2, j_p-l_p=1/2$ and $j_n-l_n=1/2,j_p-l_p=-1/2$. 
The quadrupole and spin vibration coupling strengths are defined as
\begin{equation}
\begin{split}
\alpha _2(j_p,j_n) & =\alpha ^{(0)}_2|(U^2_{j_p}-V^2{j_p})(U^2_{j_n}-V^2_{j_n})| \ , \\
\alpha _1(j_p,j_n) & = \alpha ^{(0)}_1 \ , \\
\end{split}
\end{equation}
where $V^2_j$ is the occupation probability for the level $j$ and $U^2_j=1-V^2_j$.

\begin{figure}[t]
\begin{center}
\rotatebox{0}{\scalebox{0.3}[0.3]{\includegraphics{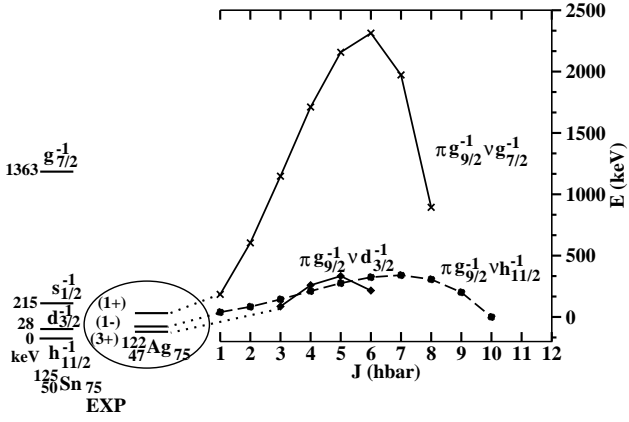}}} 
\caption[]{\label{paar}Proton-neutron quasi-particle multiplets in $^{122}$Ag. 
Neutron single-particle states in $^{125}$Sn are taken from \cite{Ka11}.}
\end{center}
\end{figure}
This approach was used extensively in the mid-shell indium isotopes \cite{Ki88}, 
where a good  overall agreement with the experimental data was achieved.
The level energies of $^{122}$Ag were calculated by using the single-particle 
neutron energies taken from the $^{125}$Sn$_{75}$ level scheme \cite{Ka11}, 
which is presented in Fig.~\ref{paar}. The neutron occupation probabilities, 
listed in Table~\ref{122specfac},  were calculated from the spectroscopic factors 
$S$, obtained in the $^{124}$Sn(d,p)$^{125}$Sn$_{75}$ reaction \cite{BH73}. The 
coupling strengths $a_1^{(0)}\approx 15/A= 0.12$ MeV and 
$a_2^{(0)}= 382 \beta _2^2 (\hbar\omega _2)^{-1} = 16.23$ MeV were deduced from 
the $\beta _2$  and the $E_{2^+}$ energy of the first phonon excitation in 
$^{124}$Cd. The proton occupation probability $V(\pi 1g_{7/2})^2=0.80$ 
is deduced from \cite{Va76}. 

\begin{table}[t]   
\caption{\label{122specfac} Spin/parities ($J^\pi$), energies ($E$), 
configurations, spectroscopic factors \cite{BH73} ($S$) and neutron occupation 
probabilities ($V_k^2$) for $^{125}$Sn$_{75}$}
\begin{center}
\begin{tabular}{cccccccccc} 
\hline\hline
$J^\pi$  & E (keV) & conf. & $S$ & $V_k^2$\\
\hline
$11/2^-$ & 0       & $\nu 1h_{11/2}$ & 0.42 & 0.58 \\
$3/2^+$  & 28      & $\nu 2d_{3/2}$  & 0.44 & 0.56\\
$1/2^+$  & 215     & $\nu 3s_{1/2}$  & 0.33 & 0.67\\
$7/2^+$  & 1363    & $\nu 1g_{7/2}$  & 0.038& 0.96\\  
\hline\hline
\end{tabular}
\end{center} 
\end{table}

Fig.~\ref{paar} shows the calculated parabolas for the 
$\pi g_{9/2}^{-1}\otimes \nu g^{-1}_{7/2}$,
$\pi g_{9/2}^{-1}\otimes \nu d^{-1}_{3/2}$ and 
$\pi g_{9/2}^{-1}\otimes \nu h^{-1}_{11/2}$ multiplets in $^{122}$Ag. 
Within each of the multiplets, the excited states decay via fast $M1$ transitions, 
and hence the observed isomeric states can arise only from transitions connecting 
states of different multiplets. The lowest-lying states, belonging to each of
the multiplets, are the $1^-$, $3^+$ and $1^+$ states, which is consistent with 
the experimental data. 

\begin{figure}[t]
\begin{center}
\rotatebox{0}{\scalebox{0.3}[0.3]{\includegraphics{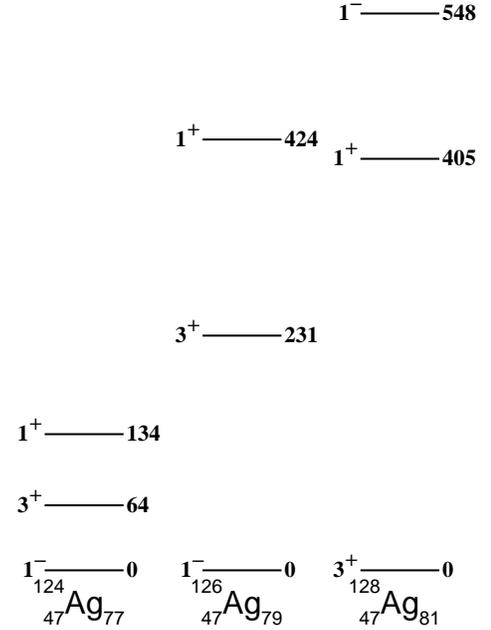}}} 
\caption[]{\label{nush126}Partial theoretical level schemes for $^{122-128}$Ag 
obtained with the jj45pna interaction.}
\end{center}
\end{figure}

Shell model calculations were also performed for $^{122-128}$Ag with the full
jj45pn space. Results are presented in Fig.~\ref{nush126}. In $^{128}$Ag, the 
main component of the $1^+$ wave function (w.f.) is 
$\pi g_{9/2}^{-1}\otimes \nu g_{7/2}^{-1}$ with an amplitude of 30\%, followed 
by $\pi p_{1/2}^{-2}\otimes \nu g_{7/2}^{-1}$ with 10\%. 
The w.f. of the $3^+$ state has a leading $\pi g_{7/2}^{-1}\otimes \nu d_{3/2}^{-1}$ 
component with an amplitude of 39\%. Weaker components are 
$\pi p_{1/2}^{-2}\otimes \nu d_{3/2}^{-1}$ and 
$\pi p_{3/2}^{-2}\otimes \nu d_{3/2}^{-1}$ with 13\% and 11\%, respectively.
The $1^-$ state consists of $\pi g_{9/2}^{-1}\otimes \nu h_{11/2}^{-1}$, 
contributing 34\% of the w.f., and $\pi p_{1/2}^{-2}\otimes \nu h_{11/2}^{-1}$ 
which has an amplitude of 24\% of the w.f. All other components contribute with 
smaller amplitudes. In the lighter nuclei, these configurations become diluted 
and the wave functions are much more fragmented. The relative position of the 
theoretical $1^+$ and $3^+$ is the same as the ordering of the experimental 
levels. However, in contrast to the spin/parity assignments made in the present 
work, the theoretical $1^-$ is pushed below the $3^+$ states in $^{124,126}$Ag. 
The behavior of the $1^-$ level in the neutron-rich even-even Ag nuclei is not 
surprising, since in the neutron-rich even-A 
indium nuclei a strong variation of the $1^-$ state with respect to the $3^+$ 
state is also observed and attributed to a weakening of the effects of the $p-n$ 
interaction when moving from $^{129}$In$_{82}$ to $^{126}$In$_{78}$ \cite{Sc04}.
It was suggested also that the underestimation of the $p-n$ interaction in the
region of nuclei with $N<82$ may limit the predictive power of the Shell model.

\section{Conclusions}
This work presents results on two previously known sub-microsecond isomers
and on two new isomeric states in $^{123,125}$Ag. Because of the high efficiency 
and granularity of RISING, coincidences were established enabling the 
construction of the level schemes. A new microsecond isomer was observed in 
$^{122}$Ag. The isomeric state in $^{124}$Ag is observed to decay via a 75-keV
transition and a half-life of the intermediate state was deduced. An isomeric 
state with a half-life of 27(6) $\mu$s is observed in $^{126}$Ag. The 
spin/parity assignments are based on the systematics and the observed decay 
pattern.

The positive-parity states of the odd-mass nuclei were analyzed within a 
$\pi g_{9/2}^{-3}$ coupling scheme, based on a simple angular momentum 
re-coupling algebra, and the results compared to shell model calculations, performed in a 
larger space. A reasonable description of the isomeric $17/2^-$ and the $13/2^-$ 
level energies is achieved. The biggest discrepancy between the theoretical 
calculations and the $^{123}$Ag data is in the $7/2^+$ level energy, which 
appears 635 keV above the ground state. In $^{123,125}$Ag, the theoretical level 
energy for the $11/2^+$ is also overestimated by approximately 300 keV. Further 
qualitative analysis was made in the framework of the cluster-vibration model.

The excited states in the even-mass silver nuclei were analyzed with the shell 
model and within a phenomenological approach based on a quadrupole and 
spin-vibrational proton-neutron interaction. 

Even though the phenomenological approaches used in the present work give a 
satisfactory description of particular levels, more realistic calculations using 
a larger valence space fail in reproducing the level energy sequence in several 
cases. This may be explained by an inapplicability of the $p-n$ interaction 
parametrized with respect to the nuclei placed close to $^{132}$Sn and by a
certain degree of collectivity presented in the nuclei with few valence holes to
the $N=82$ and $Z=50$ magic numbers. Nevertheless, the intruder $\pi g_{9/2}$ and 
$\nu h_{11/2}$ orbits seem to play an important role in understanding the 
structure of the neutron-rich silver nuclei and the isomerism in $^{122-126}$Ag 
in particular.

\acknowledgements
This work is supported by the Bulgarian National Science Fund under contract No: 
DMU02/1, UK STFC, Royal Society, the Spanish Ministerio de Ciencia e Innovaci\'on under
contracts FPA2009-13377-C02-02 and FPA2011-29854-C04-01 and the Spanish Consolider-Ingenio 2010 
Programme CPAN (CSD2007-00042), the Swedish Science Council and OTKA contract number K100835.

\end{document}